\documentclass[aps,twocolumn,nofootinbib,showpacs,prd]{revtex4}
\usepackage{graphicx}
\usepackage{amsfonts}
\usepackage{amssymb}
\usepackage{amsbsy}
\usepackage{amsmath}
\usepackage{latexsym}
\usepackage{natbib}
\usepackage{bm}
\usepackage{color}
\usepackage{float}

\begin{document}
\title{A Cosmological Reconstruction of the Higgs Vacuum Expectation Value}
\author{Soumya Chakrabarti\footnote{soumya.chakrabarti@vit.ac.in}, Anagha V\footnote{anagha.v2021@vitstudent.ac.in}, Selva Ganesh\footnote{selvaganesh.2021@vitstudent.ac.in} and Vivek Menon\footnote{vivekmenon.k2021@vitstudent.ac.in}}
\affiliation{Vellore Institute of Technology \\ 
Tiruvalam Rd, Katpadi, Vellore, Tamil Nadu 632014 \\
India}


\date{\today}

\begin{abstract}
We present a simple toy model of cosmic acceleration driven purely by a self-interacting scalar field embedded in theory of grand unification. The scalar self-interaction is Higgs-like and provokes a spontaneous symmetry breaking. The coefficient of the quadratic term in the self-interaction potential has an evolution and it leads to a cosmic variation of proton-to-electron mass ratio, $\mu$. We perform a cosmological reconstruction from the kinematic parameter jerk and discuss a few cosmological consequences of the theory. We also compare the theoretically calculated $\mu$ variation with the observations of molecular absorption spectra from Cesium Atomic Clock data.       
\end{abstract}

\maketitle

\section{Introduction} \label{intro1}
Fundamental forces are the building blocks of physics and our interpretation of the nature. In general, the laws governing these force are well understood. However, in certain cases there are contradictions in the usual chain of arguments behind standard scientific principles. Among the four fundamental forces, gravity is the most interesting and at the same time, inspires the most popular of the contradictions. Although a \textit{perfect} theory of gravity is not fully developed, one must acknowledge that the General Theory of Relativity (GR) works as the best description till date. GR introduces a correlation between kinematics of an object and the geometry of a space-time continuum. It provides explanations for astrophysical phenomena quite accurately but also leaves behind a few serious questions. One such question is related to the origin and time-history of our universe. The known universe has been expanding with acceleration/deceleration since the big bang, as a function of cosmic time \cite{riess1, beto} and it does not go well with the usual logics following from the idea of an attractive gravitational interaction alone. The most effective approach to counter-argue in this situation is to assign an additional \textit{`unknown'} energy density component to take care of this expansion. This component must have an effective negative pressure and such components are now well-apprehended as \textit{Dark Energy (DE)} fluid \cite{maor1, maor2, upadhye}. Inspite of this concept being widely discussed and re-visited for quite a few decades, no consensus has been reached over its unique identity. There are simple toy models which utilize a cosmological constant to play the role of DE. Further generalizations are written using a Quintessence scalar field \cite{zlatev, sahni, copeland} which inspire the idea of a \textit{`dynamic'} DE. These vintage approaches are mostly redundant at present due to stern challenges from the observations related to equivalence principle \cite{paddy, riess2, eisen, sen, adel}. In principle, dynamic DE models are more realistic since they can describe a smooth transition of the universe from one epoch into another. Using the evolving equation of state (EOS) for the DE in these models one can investigate if the distribution can cluster below the Hubble scale. A carefully chosen time evolving scalar field with/without an interaction potential can describe this dynamic EOS as a function of cosmic time or redshift. \\

Scalar fields carry additional motivations from particle physics and an envisaged theory of grand unification. In a theory of unification, scalar fields either come into the framework as a correction from quantum field theory or as a mass generator for fundamental particles \cite{wet, carrol}. These scalar fields are also useful for string compactification in a higher dimensional spacetime. Motivated by these pre-dominating notes, we imagine that the cosmic scalar field is embedded within the standard model and assign a non-trivial, Higgs-like self-interaction to the field. The self-interaction form leads to a cosmic variation of Higgs vacuum expectation value (VEV) as the mass term has a mild evolution \cite{camp, cal, langa, olive, dine} 
\begin{equation}
V(\phi) = V_{0} + M(t) \phi^{2} + \frac{\lambda}{4} \phi^{4}.
\end{equation}

Utilization of a self-interaction potential written in the above form is not entirely new and has been considered by theoretical physicists from time to time (see for instance \cite{sola, sc}). We note that the function \textit{$M(t)$} can be imagined as a function of cosmic time or redshift and can be correlated to a variation of quark mass through Higgs VEV. In very brief, we discuss the theoretical origin of this correlation in the next section. In a cosmological model, the non-trivial scalar self-interaction leads to some modifications of the standard Friedmann-Robertson-Walker (FRW) equations. In section $3$, we discuss the consequences of a kinematical reconstruction of this modified thoery using the cosmological jerk parameter. We conclude the article in Section $4$.

\section{A Variation of Higgs Vacuum Expectation Value}
A variation of Higgs VEV is theoretically realized through some careful tweaks in the standard model of particle physics. It can be directly measured through the proton-to-electron mass ratio, $\mu$. This variation can affect the nature of fundamental concepts related to light, electricity and magnetism, within a unified formalism known as the electroweak theory of interactions. The theory uses Higgs boson as a mass generator of elementary particles, through
\begin{equation}
m_{e,q} = \lambda_{e,q} \nu,
\end{equation}
where $m_{e,q}$ is the quark mass, $\nu$ is the VEV and $\lambda_{e,q}$ is the Yukawa coupling constant \cite{calmet1, gasser, ji, yang}. While the quark masses are directly proportional to Higgs VEV, mass of a proton $m_{p}$ depends mainly on quark-gluon interaction. The VEV contribution to $m_{p}$ can be rendered negligible and as a consequence, any pre-assigned variation in VEV affects the electron masses alone \cite{bag, casadio}. This effect can be observed directly by studying the cosmic variation (as a function of cosmic time or redshift) of the proton-to-electron mass ratio $\mu$, writing  
\begin{eqnarray}
&& \frac{\Delta \mu}{\mu} = \frac{\Delta m_{p}}{m_{p}} - \frac{\Delta m_{e}}{m_{e}} =  - \frac{91}{100} \frac{\Delta \nu}{\nu},  \label{2-10}\\&&
\Delta \nu / \nu = \frac{(\nu_{z}-\nu_{0})}{\nu_{0}}.
\end{eqnarray}
$\Delta \nu / \nu$ is the dimension-less variation of Higgs VEV as a function of redshift $z$. $\nu_{0} \sim 246 GeV$ can be used as the present value of Higgs VEV in natural units. \\

The fact that proton-to-electorn mass ratio $\mu$ can be treated as a measurable entity in the spectroscopic analysis of Quasar spectra, makes this correlation more interesting. Optical atomic clocks are utilized in these spectroscopic studies to measure any deviation from the standard model of particle physics. For the purpose of this article, we consider an observational bound on $\frac{\Delta{\mu}}{\mu}$ provided by the analysis of cesium atomic clock \cite{hunte}, written as 
\begin{equation}
\frac{\Delta{\mu}}{\mu} = (-0.5 \pm 1.6)\times 10^{-16} \, \, year^{-1} .
 \label{017}
\end{equation}

It is convenient to write the above quantity using the present value of Hubble constant $H_{0} \simeq 7 \times 10^{-11}  \, \, year^{-1}$ as a unit. Then the scale of variation become $\frac{\Delta{\nu}}{\nu} \simeq 10^{-6} H_{0}$. We will compare the theoretical variation of Higgs VEV derived from the field euations with the observational data Table. \ref{table1}.

\begin{table}
\caption{{\small $\Delta \mu / \mu$ as a function of redshift : weighted average of data from Hydrogen and other Molecular Spectra of Quasars}}\label{table1}
\begin{tabular*}{\columnwidth}{@{\extracolsep{\fill}}lrrrrl@{}}
\hline
\multicolumn{1}{c}{Quasar} & \multicolumn{1}{c}{Redshift} & \multicolumn{1}{c}{$\Delta \mu / \mu \, \, [10^{-6}]$} \\
\hline
B0218+357 & 0.685 & $ -0.35  \pm 0.12 $ \\
PKS1830-211 & 0.89 & $ 0.08  \pm 0.47 $ \\
HE0027-1836 & 2.40 & $ -7.6  \pm 10.2 $ \\
Q0347-383 & 3.02 & $ 5.1  \pm 4.5 $ \\
Q0405-443 & 2.59 & $ 7.5  \pm 5.3 $ \\
Q0528-250 & 2.81 & $ -0.5  \pm 2.7 $ \\
B0642-5038 & 2.66 & $ 10.3  \pm 4.6 $ \\
J1237+064 & 2.69 & $ -5.4  \pm 7.2 $ \\
J1443+2724 & 4.22 & $ -9.5  \pm 7.5 $ \\
J2123-005 & 2.05 & $ 7.6  \pm 3.5 $ \\
\hline
\end{tabular*}
\end{table}

The data is suggestive of a mild variation of $\mu$ during a matter-dominated deceleration ($z > 1$) as well as the subsequent late-time acceleration (near $z \sim 0$). However, no concrete claim can be made as long as we do not solve for the mathematical form of this variation. We assign this variation to a one dimensional, massive scalar field in reduced Planck Mass unit, minimally coupled to gravity. We explore the dynamics of this scalar field using the Einstein field equations. The scalar self-interaction is Higgs-like but enjoys a scope of spontaneous symmetry breaking.
\begin{equation}\label{potential}
V(\phi) = V_{0} + M(t) \phi^{2} + \lambda \phi^{4}.
\end{equation}

In the above potential $\lambda$ is a dimension-less number, treated as a parameter of the theory. The term $M(t)$ works like an effective time-evolving coupling, correlated to the $W-$boson mass measurements from collider experiments through the Higgs VEV $\nu$. We can calculate $\nu$ from the potential as
\begin{eqnarray}
&& \frac{\partial V}{\partial \phi} \bigg\rvert_{\nu} = 0,
\nu = \sqrt{\frac{-2M(t)}{\lambda}}.  \label{328}
\end{eqnarray}

We also find it useful to write any variation in $\nu$ as $\nu(z)$, i.e., a function of redshift. We enforce this by introducing $M(t) = M_{0} M(z)$ and assigning the present observational bound on the W-boson mass to $M_{0}$. Thereafter, $M(z)$ can evolve as a dimension-less function and contribute to the proton-to-electron mass ratio. \\

We assume no additional matter lagrangian term $L_{m}$ and portray the cosmological dynamics to be scalar field dominated. The scalar field energy-momentum contribution is found to be
\begin{equation}\label{minimallyscalar}
T^\phi_{\mu\nu} = \partial_\mu\phi\partial_\nu\phi - g_{\mu\nu}\Bigg[\frac{1}{2}g^{\alpha\beta}\partial_\alpha\phi\partial_\beta\phi - V(\phi)\Bigg]. 
\end{equation}

\section{Reconstruction from Jerk Parameter}
Having established a reasonable setup to describe $\mu$-variation cosmologically, we discuss whether or not we can fit such a cosmological model with the very basic observational data-sets describing late-time acceleration of our universe. We assume that the expanding universe is homogeneous and isotropic and describe it by the Friedmann-Robertson-Walker (FRW) metric,
\begin{equation}
ds^{2} = -dt^{2} + a^{2}(t)\left(dr^{2} + r^{2}d\Omega^{2}\right). \label{eq:frwmetric}
\end{equation}
The cosmological scale factor $a(t)$ and the scalar field $\phi(t)$ are spatially homogeneous. From this point onwards we work in a natural unit system ($8 \pi G = 1$) and express cosmic time-derivative with a dot. The Einstein field equations are derived as a set of coupled ordinary differential equations,
\begin{equation} \label{fe1minimal}
3\Big(\frac{\dot{a}}{a}\Big)^{2} = \rho_{\phi} = \frac{\dot{\phi}^{2}}{2} + V\left( \phi \right),
\end{equation} 
 
\begin{equation} \label{fe2minimal}
-2\frac{\ddot{a}}{a}-\Big(\frac{\dot{a}}{a}\Big)^{2} = p_{\phi} = \frac{\dot{\phi}^{2}}{2}-V\left( \phi \right),
\end{equation}
and
\begin{equation} \label{phiminimal}
\ddot{\phi} + 3\frac{\dot{a}}{a}\dot{\phi} + \frac{dV(\phi)}{d\phi} = 0.  
\end{equation}     

The last equation is derived from a $\phi$-variation of the action, however, this is not an independent equation. A straightforward way is to try and find an exact solution of the above set. However, it is also apparent that such a solution is not straightforward to find out and if at all possible, a solution can only be extracted for special cases. We follow a route of \textit{reverse engineering} and find a solution of the field equations starting from a kinematic parameter. A properly chosen evolution of kinematic parameter ensures the correct dynamics required for a present phase of acceleration, preceded by an epoch of matter-dominated deceleration \cite{paddy}. While one can not clearly identify the nature of matter constituents of the universe driving this accelerated/decelerated expansion, the kinematics can be studied from the measurement of Hubble-free Luminosity distance measurement of Type-IA Supernova \cite{beto}. The best theoretical model which can fit in with the kinematics of these expanding phases and describe a smooth (no discontinuity) transition from deceleration into acceleration, is the $\Lambda$CDM model. A $\Lambda$CDM model is usually classified by a characteristic \textit{jerk} parameter value $j = 1$, which is also widely accepted as the present value of jerk. Jerk is defined as the third order kinematic parameter related to the cosmological scale factor, whereas the first and second order parameters are Hubble $H$ and deceleration parameter $q$. 

\begin{eqnarray}
&& H = \frac{\dot{a}}{a} ~;~ q = -\frac{\ddot{a}a}{\dot{a}^2} = -\frac{\dot H}{H^2}-1, \\&&
j = \frac{\stackrel{\bf{...}}{a}}{aH^3} = \frac{\ddot H}{H^3}+3\frac{\dot H}{H^2} + 1 \label{eq:eq1.3}.
\end{eqnarray}

If $j = 1$, the definition of jerk simply provides a differential equation which can be solved to write
\begin{equation} \label{eq8}
a\left(t\right) = \left(Ae^{\alpha t} + B e^{-\alpha t}\right)^{\frac{2}{3}}.
\end{equation}

\begin{figure}[t!]
\begin{center}
\includegraphics[width=\linewidth]{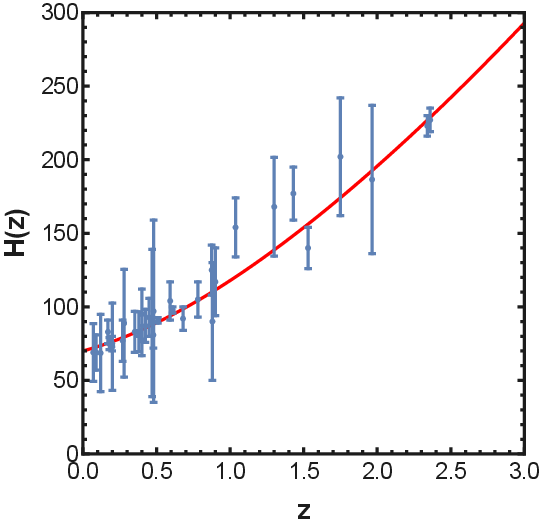}
\includegraphics[width=\linewidth]{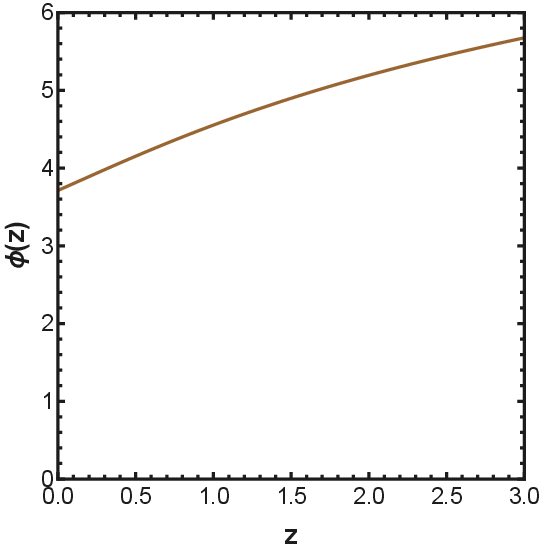}
\caption{Profile of the reconstructed Hubble function $H(z)$ and the scalar field $\phi(z)$ : Model $1$}
\label{Model_1_graph_1}
\end{center}
\end{figure}

$A$ and $B$ are constants coming out of the integration and treated as parameters of the theory. We use the usual re-scaling $a = 1/(1+z)$ (where the present value of the scale factor is scaled to $1$, realized at $z = 0$) in Eq. (\ref{eq8}) and write
\begin{equation}
H\left(z\right)=\frac{2}{3}\lambda\left[1 - 4AB\left(1+z\right)^3\right]^{\frac{1}{2}}.
\end{equation}
At $z = 0$, we assign $H(z)|_{z=0} \equiv H_{0} = \frac{2}{3}\lambda\left(1 - 4AB^{}\right)^{\frac{1}{2}}$ and express the Hubble function as 
\begin{equation} \label{eq13}
H\left(z\right)=\frac{H_0}{\left(1-4AB\right)^{\frac{1}{2}}}\left[1-4AB\left(1+z\right)^3\right]^{\frac{1}{2}}.
\end{equation} 

Eq. (\ref{eq13}) is the description of Hubble as a function of look-back time. The observed present value of $H_0$ is $\sim 70(km/s)/Mpc$ which enables us to write Eq. (\ref{eq13}) as
\begin{equation} \label{eq14}
H\left(z\right) = \frac{70}{\left(1 - x_0\right)^{\frac{1}{2}}}\left[1 - x_0\left(1+z\right)^3\right]^{\frac{1}{2}}.
\end{equation}
For convenience, we have wrtten $4AB = x_0$. Eq. (\ref{eq14}) describes the standard $\Lambda$CDM cosmological model and we shall refer to this as Model-$I$. \\

Recent research works on the distribution of dark energy and the resulting Hubble tension collectively suggest that a $\Lambda$CDM model may not be \textit{the unique} model to describe the present acceleration. In fact, in some articles it has been pointed out that one can model a reasonable desription of the late-time cosmic expansion with a small departure from $\Lambda$CDM \cite{sc} as well. To cross-check this, we write a simple modified toy model by introducing a new parameter $\delta$ in Eq. (\ref{eq14}) as follows
\begin{equation} \label{eq15}
H\left(z\right) = \frac{70}{\left(1 - x_0\right)^{\frac{1}{2}}}\left(1 - x_0\left(1+z\right)^3\right)^{\frac{1}{2} + \delta}.
\end{equation}
This describes a modified model of cosmic expansion as one can easily check by calculating the jerk parameter. We shall refer to this as Model-$II$. \\

\begin{figure}[t!]
\begin{center}
\includegraphics[width=\linewidth]{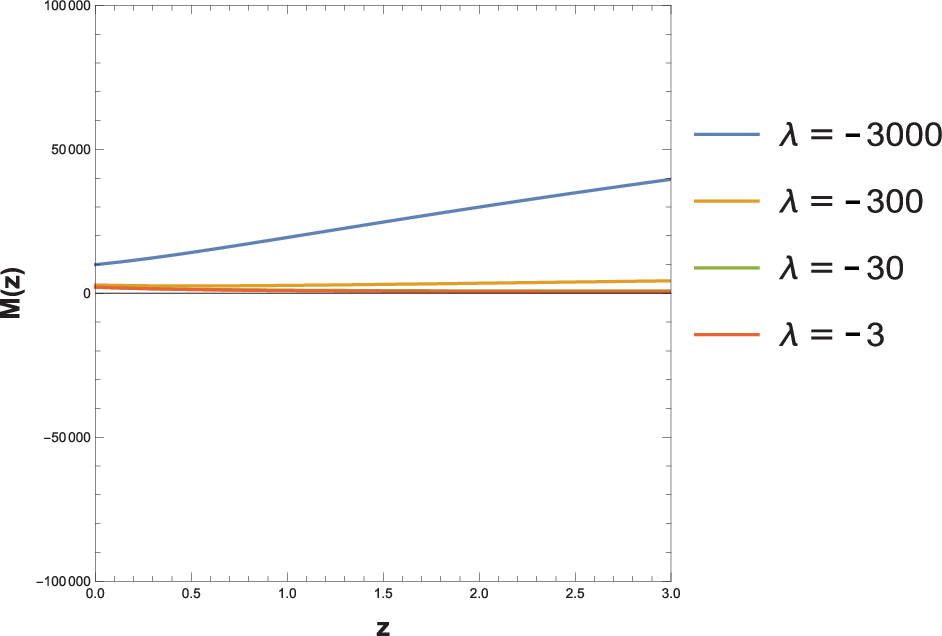}
\includegraphics[width=\linewidth]{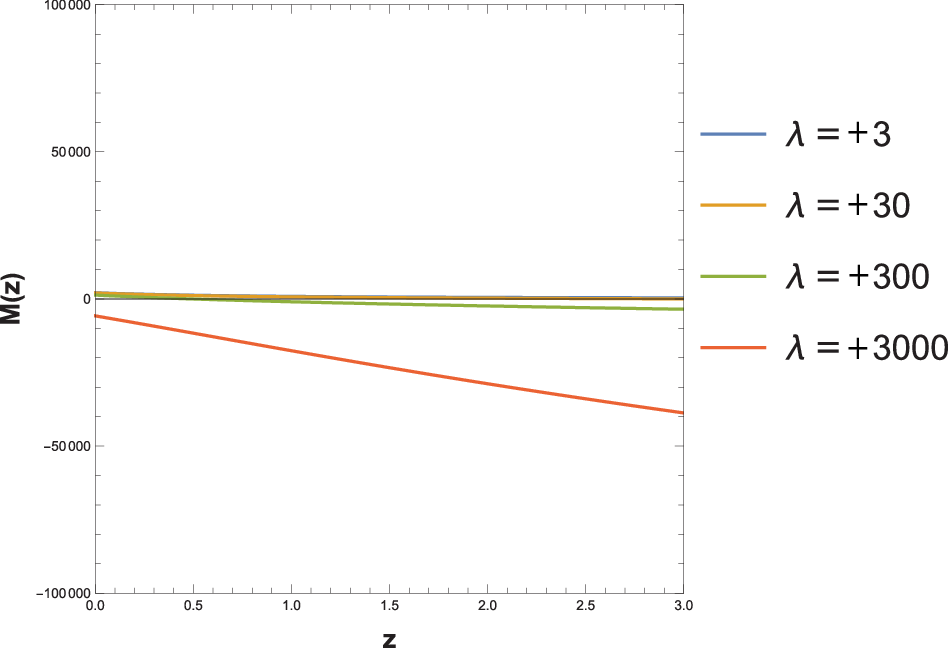}
\caption{Profile of the mass function $M(z)$ : Model $1$}
\label{Model_1_graph_2}
\end{center}
\end{figure}

\begin{figure}[t!]
\begin{center}
\includegraphics[width=\linewidth]{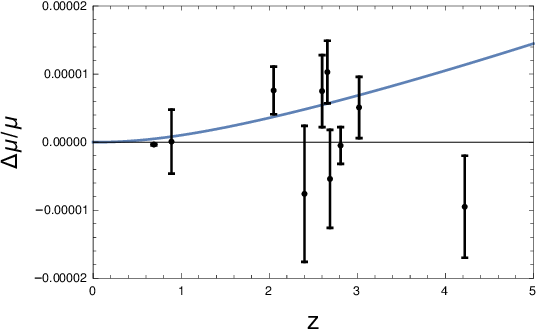}
\caption{Evolution of proton to electron mass ratio $\frac{\Delta \mu}{\mu}$ as  afunction of redshift : Model $1$}
\label{Model_1_graph_3}
\end{center}
\end{figure}

We use Eq. (\ref{eq15}) to plot $H(z)$ v/s $z$ for two different values of $x_0 \space (-0.355 and -0.300)$, each with $\delta = 0.0001$. We also show the observational data points showing the measured values of Hubble as a function of $z$ alongwith respective error bars. Furthermore, we solve the two field Eqs. (\ref{fe1minimal}) and (\ref{fe2minimal}) together to write a correlation between the scalar field and the Hubble function as
\begin{equation} \label{eq17}
\dot{\phi}^2 = -2\dot{H}.
\end{equation}

We convert the time derivative into a derivative of red-shift and write
\begin{equation} \label{eq18}
\left(\frac{\mathrm{d}\phi}{\mathrm{d}z}\right)^2 = 2\frac{\frac{\mathrm{d}H}{\mathrm{d}z}}{(1+z)H(z)}
\end{equation}

For the chosen parameter values, we solve Eq. (\ref{eq18}) to find the evolutions of the scalar field $\phi$ and plot them as a function of $z$. The evolving coupling connected to Higgs VEV, $M(z)$ was introduced as part of the scalar interaction $V(\phi)$, through $V{\left(\phi\right)} = V_0 + M{\phi}^2 + {\lambda{\phi}^4}$.

\begin{figure}[t!]
\begin{center}
\includegraphics[width=\linewidth]{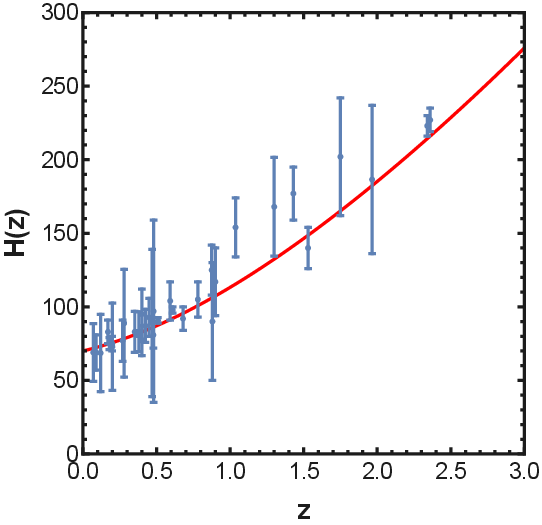}
\includegraphics[width=\linewidth]{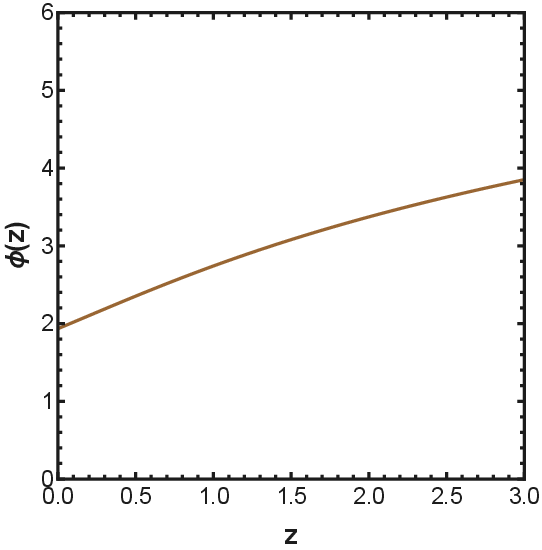}
\caption{Profile of the Hubble function $H(z)$ and the scalar field $\phi(z)$ : Model $2$}
\label{Model_2_graph_1}
\end{center}
\end{figure}

Using the field equations we solve for $M$ and write

\begin{equation} \label{eq20}
M(\phi) = \frac{3H^2 - \frac{1}{2}\dot{\phi}^2 - \mathrm{V_0} - \lambda\phi^4 }{\phi^2}.
\end{equation} 
$V_0$ and $\lambda$ are constant parameters and their choices determine the profile of $M(\phi)$ and in turn, the Higgs VEV. Since we know the profile of Hubble and the scalar field as a function of $z$, we can plot the evolution of $M(\phi)$ (or $M(z)$) for a wide range of values of $\lambda$ and $V_0$. In Fig. \ref{Model_1_graph_1} we plot the evolution of Hubble function and the scalar field as a function of redshift for Model $I$. It can be seen that the Hubble function profile is quite consistent with observational data-points until $z \sim 2.5$. This range involves the matter-dominated deceleration and the subsequent late-time cosmic acceleration. The solution for scalar field is also portrayed and it shows a monotonically increasing profile. Using Eq. (\ref{328}), we derive the expressions for $M(z)$ and plot in Fig. \ref{Model_1_graph_2}. The evolution of $M(z)$ depends on the choice of $\lambda$ which is evident from Eq. (\ref{eq20}). However, we can easily rule out half of the cases to avoid unphysical outcomes as in a negative mass profile. Finally, using Eq. (\ref{2-10}), we calculate the change in proton-to-electron mass ratio $\frac{\Delta \mu}{\mu}$ and plot it in Fig. \ref{Model_1_graph_3}, alongwith the data-points from Quasar absorption spectra as in Table \ref{table1}. It shows a satisfactory fit in between theory and observations. \\   

\begin{figure}[t!]
\begin{center}
\includegraphics[width=\linewidth]{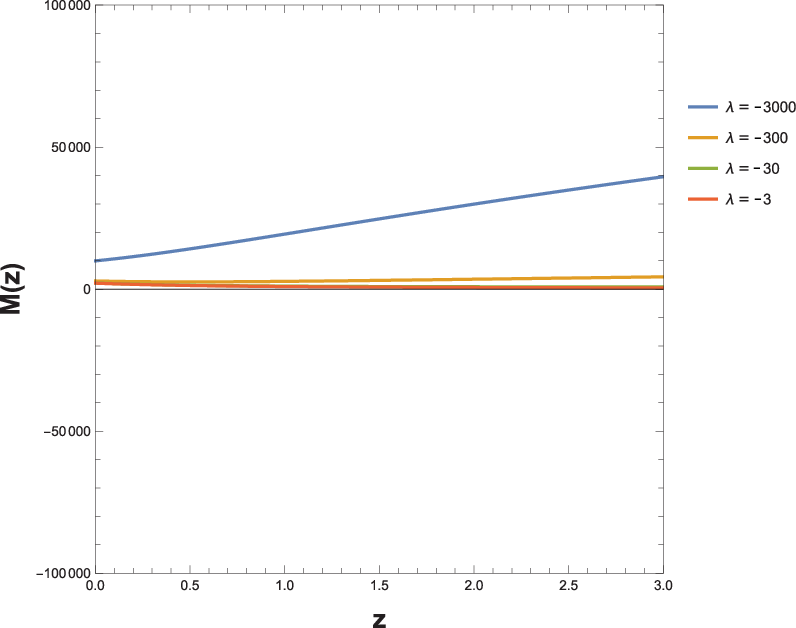}
\includegraphics[width=\linewidth]{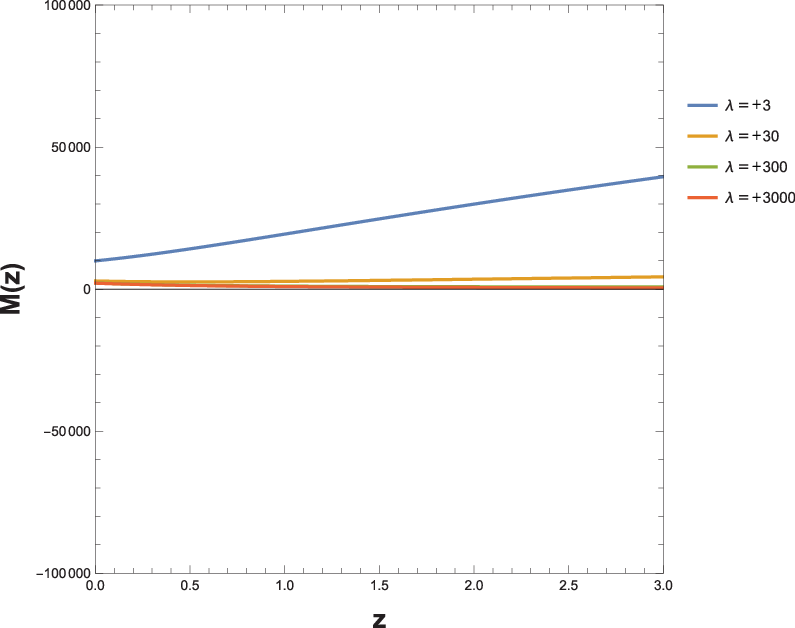}
\caption{Profile of the mass function $M(z)$ : Model $2$}
\label{Model_2_graph_2}
\end{center}
\end{figure}

\begin{figure}[t!]
\begin{center}
\includegraphics[width=\linewidth]{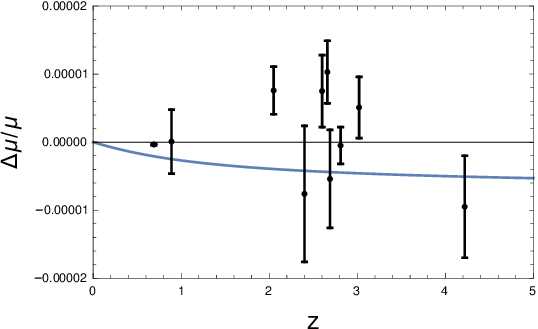}
\caption{Evolution of proton to electron mass ratio $\frac{\Delta \mu}{\mu}$ as  afunction of redshift : Model $2$}
\label{Model_2_graph_3}
\end{center}
\end{figure}

Similarly, in Fig. \ref{Model_2_graph_1} we plot the evolution of Hubble function and the scalar field as a function of redshift for Model $II$. We see that the Hubble function profile is once again consistent with observational data-points. The solution for scalar field is also described and it shows a monotonically increasing profile. Using Eq. (\ref{328}), we derive $M(z)$ and plot it in Fig. \ref{Model_2_graph_2}. $M(z)$ depends on the choice of $\lambda$ and in this case, for positive as well as for negative values of $\lambda$, we can find realistic (positive) mass profiles. Finally, using Eq. (\ref{2-10}) once again, we calculate the change in proton-to-electron mass ratio $\frac{\Delta \mu}{\mu}$ and plot in Fig. \ref{Model_2_graph_3}, alongwith measured data-points from Table \ref{table1}. For this model too, we find a satisfactory fit in between theory and observations. \\

\begin{figure}[t!]
\begin{center}
\includegraphics[width=\linewidth]{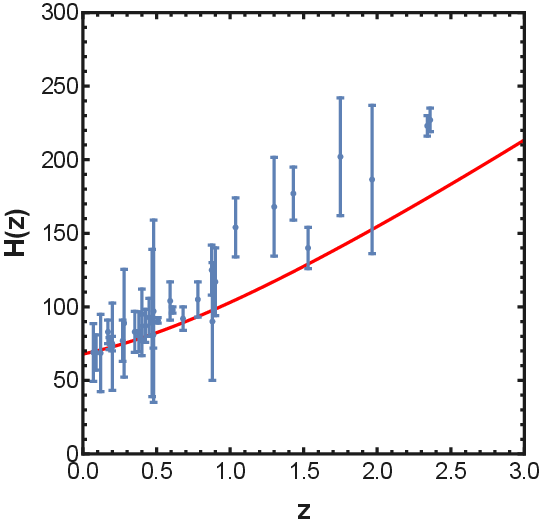}
\caption{Profile of Hubble function $H(z)$ for for a bad choice of $\delta = 0.1$}
\label{bad_Model}
\end{center}
\end{figure}

However, not all values of $\delta$ should generate a cosmology close to $\Lambda$CDM or fit in with the cosmological observations. This can easily be clarified if one studies an example with a comparatively larger choice of $\delta = 0.1$. For this case, the Hubble function is drawn in Fig. \ref{bad_Model} and the departure from observational data-points even for low redshift values is quite clear. \\

An interesting question to ask at this pont would be the following : what is the allowed range of departure from standard $\Lambda$CDM model, in terms of this departure parameter $\delta$, which is engineered within the theory. We find an answer to this question by estimating best fit values and constraints on all the model parameters (including $\delta$). We draw the confidence contours of all the parameters on the parameter space along with 2$\sigma$ error regions. We use a set of data-s found from the (i) Joint Light-Curve Analysis of $SDSS-II$ and $SNLS$ collaborations \cite{beto}, (ii) measurement of the Hubble parameter or OHD \cite{ohd, ohd1, ohd2, ohd3, ohd4, ohd5, ohd6} and BAO : Baryon Acoustic Oscillation \cite{beu, boss}. 

\begin{figure}
\begin{center}
\includegraphics[angle=0, width=0.52\textwidth]{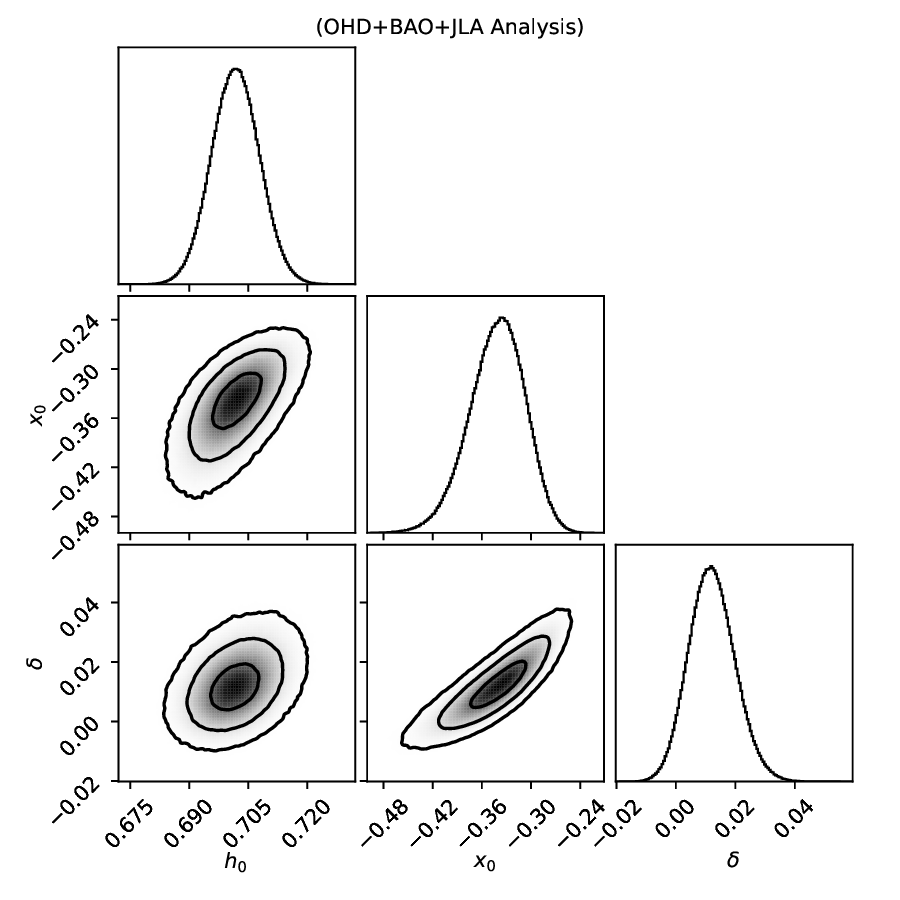}
\caption{Estimation of three parameters, confidence contours and marginalized likelihood function using OHD+JLA+BAO data.}
\label{Modelcontour}
\end{center}
\end{figure}

\begin{table}
\caption{{\small Estimated best fit values of three model parameters and the associated 1$\sigma$ uncertainty}}\label{resulttable}
\begin{tabular*}{\columnwidth}{@{\extracolsep{\fill}}lrrrrl@{}}
\hline
 & \multicolumn{1}{c}{$h_0$} & \multicolumn{1}{c}{$x_{0}$} & \multicolumn{1}{c}{$\delta$} \\
\hline
$OHD+JLA+BAO$ 	  & $0.702^{+0.006}_{-0.006}$ &$-0.34^{+0.032}_{-0.035}$ & $0.012^{+0.008}_{-0.007}$ &\\
\hline
\end{tabular*}
\end{table}

We use the theoretical model found in Eq. (\ref{eq15}) to estimate three parameters : (a) the present value of Hubble parameter, written in a dimensionless form $h_{0} = H_{0}/100 km\mbox{Mpc}^{-1} \mbox{sec}^{-1}$, (b) the integration constants combined and written as $x_0$ and (c) the departure parameter $\delta$. A python-based Markov Chain Monte Carlo simulation is used to do the statistical analysis \cite{mcmc}. We give the best fit parameter values and associated 1$\sigma$ error values in Table \ref{resulttable}. The present value of Hubble parameter is $70.2 km\mbox{Mpc}^{-1} \mbox{sec}^{-1}$ which is consistent with recent observations \cite{ohd5}. The departure parameter $\delta$ is estimated to be within the range $\sim 0.012^{+0.008}_{-0.007}$ which clearly suggests that a model of late-time cosmology need not necessarily be a $\Lambda$CDM. \\

\begin{figure}
\begin{center}
\includegraphics[angle=0, width=0.52\textwidth]{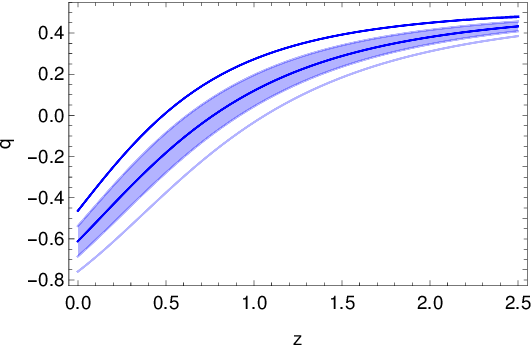}
\includegraphics[angle=0, width=0.52\textwidth]{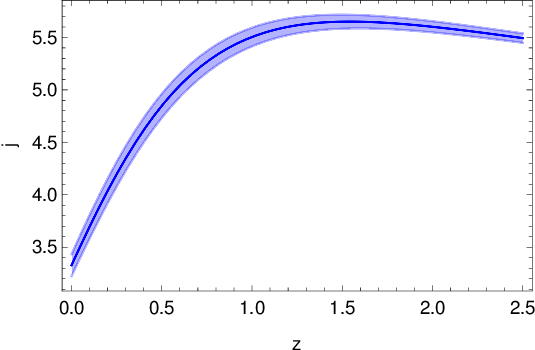}
\includegraphics[angle=0, width=0.52\textwidth]{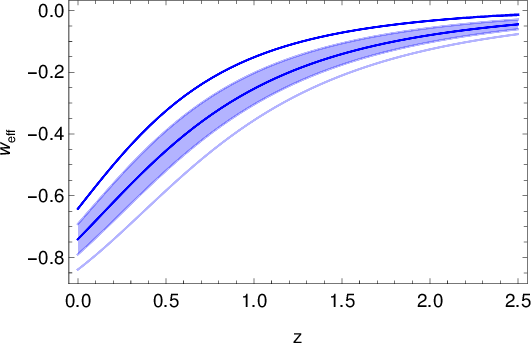}
\caption{Kinematic parameters and the effective Equation of state as a function of redshift.}
\label{qjw}
\end{center}
\end{figure}

The cosmic evolution is also understood using higher order kinematic parameters such as deceleration ($q$) and jerk ($j$). For a standard $\Lambda$CDM model these are quite well-known, however, for the modified model as in Eq. (\ref{eq15}), the evolutions are expected to be different and should be calculated. These two are drawn in Fig. \ref{qjw}, as a function of redshift, alongwith associated $1\sigma$ and $2\sigma$ error regions (shaded regions in each of the curves). The present value of deceleration parameter is around $-0.6$. Moreover, for a value of redshift $z_{t} < 1$, an evolution of the deceleration parameter from positive values into into the negative region is noted. This indicates a smooth transition of the universe from a decelerated phase into an accelerated phase of expansion. The jerk parameter is not exactly equal to $1$, signifying the fact that there is a departure from standard $\Lambda$CDM. We also calculate the effective EOS of the cosmological system as a function of redshift using the definition
\begin{equation}
w_{eff} = \frac{p_{tot}}{\rho_{tot}}.
\end{equation}
$\rho_{tot}$ and $p_{tot}$ are regarded as the total energy density and the total effective pressure. Both of these quantities can be determined from the field equations of the theory. Therefore the effective EOS is directly connected to the Hubble expansion rate by the following relations
\begin{eqnarray}
&&\frac{\rho_{tot}}{\rho_{c0}}=\frac{H^2(z)}{H^2_0},\\&&
\frac{p_{tot}}{\rho_{c0}}=-\frac{H^2(z)}{H^2_0}+\frac{2}{3}\frac{(1+z)H(z)H'(z)}{H^2_0}.
\end{eqnarray}

In the above expressions, $\rho_{c0} = 3H^2_0/8\pi G$ is called the critical density at $z = 0$. The plot of effective EOS in Fig. \ref{qjw} clearly indicates a dark energy dominated ($w_{eff} \rightarrow -1$) acceleration around $z \sim 0$ and a matter-dominated deceleration for $z > 1$. This result mimics a corresponding $\Lambda$CDM cosmology. \\

Before conclusion, we investigate whether the modified cosmological scenario can be regarded as thermodynamically stable. Given any cosmological system such a stability analysis can be done in comparison with laws of blackhole thermodynamics \cite{gibbons, jacobson, ray}. Much like a spherical black hole, a spatialy homogeneous spherical cosmological universe is surrounded by a \textit{Hubble horizon}. We define the total entropy as a sum of the boundary entropy term and the entropy contributions of any of the constituent elements, i.e., $S = S_f + S_h$. The standard laws of thermodynamics suggests that for an equilibrium \cite{jamil}
\begin{eqnarray}
&& \frac{dS}{dn}\geq 0, \\&&
\frac{d^2S}{dn^2} < 0. \label{therm}
\end{eqnarray}
These conditions are written in terms of a modified variable $n = \ln{a}$. One can improvize further to write them as 

\begin{eqnarray}\label{Sdn}
&& S_{,n} \propto \frac{(H_{,n})^2}{H^4}, \\&&\label{Entropy_Psi}
S_{,nn} = 2S_{,n}\left(\frac{H_{,nn}}{H_{,n}}-\frac{2H_{,n}}{H}\right) = 2S_{,n}\Psi.
\end{eqnarray}

\begin{figure}
\begin{center}
\includegraphics[width=0.40\textwidth]{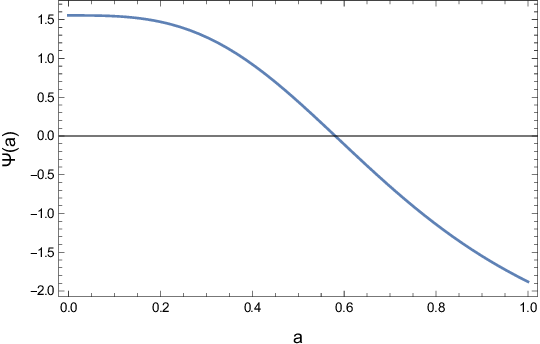}
\caption{Evolution of $\left(\frac{H_{,nn}}{H_{,n}}-\frac{2H_{,n}}{H}\right) = \Psi$ as a function of $a$ for the best fit parameter values of $h_{0}$, $x_{0}$ and $\delta$.}
\label{psiplot}
\end{center}
\end{figure}

For a thermodynamic equilibrium one needs $S_{,nn} < 0$. From Eq. (\ref{Entropy_Psi}) it is straighforward to deduct that $\Psi < 0$ is the required condition an equilibrium. We plot $\Psi$ for the modified model in Fig. \ref{psiplot}. The function is positive during an early expansion of the universe (small $a$), however, evolves into negative values once a late time acceleration starts. It can also be checked that this evolution follows a corresponding $\Lambda$CDM behavior very closely \cite{sc}.

In summary, we note that it is always possible to reconstruct the cosmic expansion history simply from any one of the kinematic parameters, provided we are able to solve the corresponding higher order differential equation. Whether or not this expansion history agrees with astrophysical observation can also be adjudged using statistical methods. The analysis presented in this article is not exhaustive. However, it helps us define the basic requirements for a highly sought after cosmological viability. Moreover, in this article the reconstruction scheme assists in modelling a cosmic acceleration being driven solely by a spontaneous symmetry breaking scalar field with a field dependent mass term. This pattern of scalar field is envisaged in a theory of Grand unification, however, to reverse engineer its profile out of a cosmological reconstruction is the essence of this article. One can cross-check using Eq. (\ref{eq20}) that the analysis only works as long as $M(\phi)$ is a function. The moment we treat this as a constant, Eq. (\ref{eq20}) provides a very steep constraint.  

\section{Conclusion}
The article presents a new model of cosmic expansion driven purely by a spontaneous symmetry breaking scalar field. The participants in this story are mathematical entities, namely, a spherically symmetric, spatially homogeneous metric tensor describing expansion and a scalar field, driving this expansion. Our motivation is to use the interplay between these two components and find evidence for an envisaged unification of gravity and physics of interacting particles. The scalar field in this model provokes a variation in the vacuum expectation value and generates a measurable evolution in the proton-to-electron mass ratio, $\mu$. In this article we have asked a very simple question : if this scalar field configuration, on its own, (without any other source of matter) is enough to drive the cosmic expansion? The answer is affirmative. \\

We reverse-engineer the field equations of this theory, starting from one of the most well-known kinematic parameters, namely, the jerk. Through this reconstruction we find a cosmological scale factor. The scale factor can describe an expanding universe alongwith the smooth transition into late-time acceleration from a preceding deceleration. The scalar field can drive this expansion and transition consistently, if and only if a variation of the Higgs VEV is allowed. From this variation we calculate the profile of $\mu$ variation and match it with the data-points found from molecular absorption spectra of quasars. We do not claim that this is a unique way forward to describe the variation embedded in a theory of gravity, however, it may be a first step towards a more generalized unified formalism.  \\

A cosmological reconstruction from jerk parameter is very simple in nature as we can write the Hubble function in closed form, in terms of cosmic time as well as redshift. This allows us to match the functional forms with direct estimations of Hubble parameter for different values of redshift. We also introduce a new parametrization of Hubble function as in Eq. (\ref{eq15}), through a new \textit{departure} parameter $\delta$. We find out that this correction parameter should not be arbitrary. It has a preferable range, as we estimate using a python-based Markov Chain Monte Carlo simulation. In this process we also estimate the best fit parameter value of Hubble parameter to be $70.2 km\mbox{Mpc}^{-1} \mbox{sec}^{-1}$. Alongwith an associated 1$\sigma$ error range, the departure parameter $\delta$ is estimated to be $\sim 0.012^{+0.008}_{-0.007}$. This leads us to a general deduction that a late-time cosmology need not be described uniquely a $\Lambda$CDM model (for which $\delta = 0$). The evolution of two kinematic parameters deceleration and jerk parameter are studied. The present value of deceleration parameter and the transition point where there is a smooth transition of the universe from deceleration into acceleraton is found to be consistent with observation. A study of effective equation of state of the cosmological system as a function of redshift clearly indicates a dark energy dominated ($w_{eff} \rightarrow -1$) acceleration at present and a matter-dominated deceleration for $z > 1$.

\section*{Acknowledgement}
Dr. Soumya Chakrabarti acknowledges Vellore Institute of Technology for the financial support through its Seed Grant (No. SG20230027), year 2023.

\end{document}